\documentclass[a4paper]{jpconf}
\usepackage{mkfig}
\usepackage{graphicx}
\begin{document}
\title{{\em Ulysses} and IBEX Constraints on the Interstellar Neutral
 Helium Distribution}

\author{Brian E. Wood$^1$ and Hans-Reinhard M\"{u}ller$^2$}

\address{$^1$ Naval Research Laboratory, Space Science Division,
  Washington, DC 20375, USA}
\address{$^2$ Department of Physics and Astronomy, Dartmouth College,
  Hanover, NH 03755, USA}

\ead{brian.wood@nrl.navy.mil}

\begin{abstract}

     We relax the usual assumption of Maxwellian velocity
distributions in the interstellar medium (ISM) in the analysis of
neutral He particle data from {\em Ulysses} and the {\em Interstellar
Boundary Explorer} (IBEX).  For {\em Ulysses}, the possibility that
a narrow component from heavy neutrals is contaminating the He signal
is considered, which could potentially explain the lower
ISM temperature measured by {\em Ulysses} compared to IBEX.
The expected heavy element contribution is about an order of
magnitude too small to resolve that discrepancy.  For IBEX, we find
that modest asymmetries in the ISM velocity distribution can
potentially improve the quality of fit to the first two years of data,
and perhaps improve agreement with the {\em Ulysses} measurements.

\end{abstract}

\section{Introduction}

     The current {\em Interstellar Boundary EXplorer} (IBEX)
and now defunct {\em Ulysses} spacecraft both have observed neutral
helium from the interstellar medium (ISM) flowing through the inner
solar system, which can be used to establish the nature of the
undisturbed ISM surrounding the Sun.  Analysis of these particle data
generally involves a forward modeling procedure assuming a Maxwellian
He velocity distribution at infinity, which is then propagated into
the inner heliosphere under the influence of gravity, where the
resulting distribution can then be used to confront actual
observations.  It is possible, however, that the particle flows
encountered by {\em Ulysses} and IBEX may not arise from strictly
Maxwellian distributions, and we here report on some initial efforts
to explore whether evidence can be found for this in the data.

     We consider both {\em Ulysses} and IBEX data here, but the
nature of the non-Maxwellian behavior that we look for is very
different for the two.  For {\em Ulysses}, which unlike IBEX is not
able to distinguish between different species of atomic neutrals, we
investigate whether the primary He distribution might be contaminated
by a weak, narrower distribution from heavy elements.  Such
contamination could potentially lead to an underestimate of the ISM
temperature.  Correcting for this would therefore increase the
{\em Ulysses} temperature measurements, potentially improving
agreement with IBEX measurements (McComas et al.\ 2015).  For IBEX,
with its superior signal-to-noise, we assess the detectability of
asymmetries in the ISM velocity distribution.

\section{Heavy Element Contamination of {\em Ulysses} Data}

     The most recent analyses of IBEX data have reported
an ISM He velocity vector consistent with that measured from
{\em Ulysses} data, but with a rather high temperature of
$T=8710^{+440}_{-680}$~K (Leonard et al.\ 2015; M\"{o}bius et al.\ 2015;
McComas et al.\ 2015), though Schwadron et al.\ (2015) report
a somewhat lower $T=8000\pm 1300$~K.  These measurements are higher than
the $T=6300\pm 340$~K temperature reported by Witte (2004) from
{\em Ulysses} measurements.  A more recent analysis of {\em Ulysses}
data has revised this temperature upwards to $T=7260\pm 270$~K (Wood
et al.\ 2015a), consistent with an independent reanalysis from Bzowski
et al.\ (2014).  This is closer to the desired IBEX temperature, but
still a bit low.  Wood et al.\ (2015b) recently studied whether
consideration of heavy element contamination to the {\em Ulysses} data
could lead to a further revision upwards.

     After He, the most abundant ISM neutrals detectable by
{\em Ulysses} should be O and Ne.  The heavier weights of these
atoms will yield narrower velocity distributions than that of He.
Thus, the effect of O and Ne contamination on the observed particle
beam is to narrow it, thereby leading to an underestimate in the
temperature.  If the real ISM temperature is $T\approx 8500$~K as IBEX
suggests, the question is how much O and Ne contamination there has to
be in {\em Ulysses} data for $T=7000-7500$~K to be perceived instead,
when fitting the data assuming the signal is 100\% He, $T=7000-7500$~K
being the actual temperature measured by Bzowski et al.\ (2014) and
Wood et al.\ (2015a).

     The degree of heavy element contribution can be quantified as
\begin{equation}
R=\left( \frac{A_O}{A_{He}} \cdot \frac{S_O}{S_{He}} +
  \frac{A_{Ne}}{A_{He}} \cdot \frac{S_{Ne}}{S_{He}} \right)
  \frac{D_{ONe}}{D_{He}},
\end{equation}
where the neutral abundances of He, O, and Ne in the ISM are indicated
by $A_{He}$, $A_{O}$, and $A_{Ne}$, respectively.  Interstellar
neutrals suffer ionization as they travel through the heliosphere,
particularly photoionization by solar EUV photons, and the various $S$
parameters in the equation are the survival probabilities for He, O,
and Ne.  The $D_{ONe}/D_{He}$ factor is the ratio of detection efficiencies
of O and Ne relative to He.  We assume the energy dependence of
detection efficiency is about the same for all these elements and that
$D_{ONe}(E)$ therefore can be approximated as a simple scalar multiple
of $D_{He}(E)$ (Banaszkiewicz et al.\ 1996; Yamamura \& Tawara 1996).
Assuming $\frac{D_{ONe}}{D_{He}}\approx 3$ for the detection efficiency
ratio (Banaszkiewicz et al.\ 1996; Yamamura \& Tawara 1996),
$\frac{A_O}{A_{He}}=3.2\times 10^{-3}$ and
$\frac{A_{Ne}}{A_{He}}=3.9\times 10^{-4}$ for the ISM neutral
abundance ratios (Slavin \& Frisch 2008), and
$\frac{S_O}{S_{He}}\approx 0.20$ and $\frac{S_{Ne}}{S_{He}}\approx 0.45$
for the survival probability ratios (Bzowski et al.\ 2013) leads
to a value of $R=0.0025$ from equation (1).

     Assuming $T=8500$~K, Figure~1 shows simulated data for a
{\em Ulysses} observation from 2001~December~12.  Maps are shown for
$R=0.003$, $R=0.03$, and $R=0.1$.  The beam becomes noticably narrower
as $R$ increases, corresponding to more O and Ne contamination.
These simulated data are fitted using the same methods as described
by Wood et al.\ (2015a), not only for this particular {\em Ulysses}
observation but for a total of 20 separate {\em Ulysses} observations
from different parts of its orbit, as defined by an orbital phase
$\phi$, where $\phi=0.5$ corresponds to the ecliptic plane crossing
near solar perihelion (see Wood et al.\ 2015a).  The results are
shown in Figure~2.  For low $R$ the actual $T=8500$~K value is
recovered, as there is insufficient contamination from O and Ne to
affect the synthetic data, but $T$ starts to lower significantly
for $R\geq 0.003$.  By $R\approx 0.03$, temperatures consistent
with the measurements of the real {\em Ulysses} data are
being observed, with some orbital phase dependence being apparent.
\begin{figure}[p]
\plotfiddle{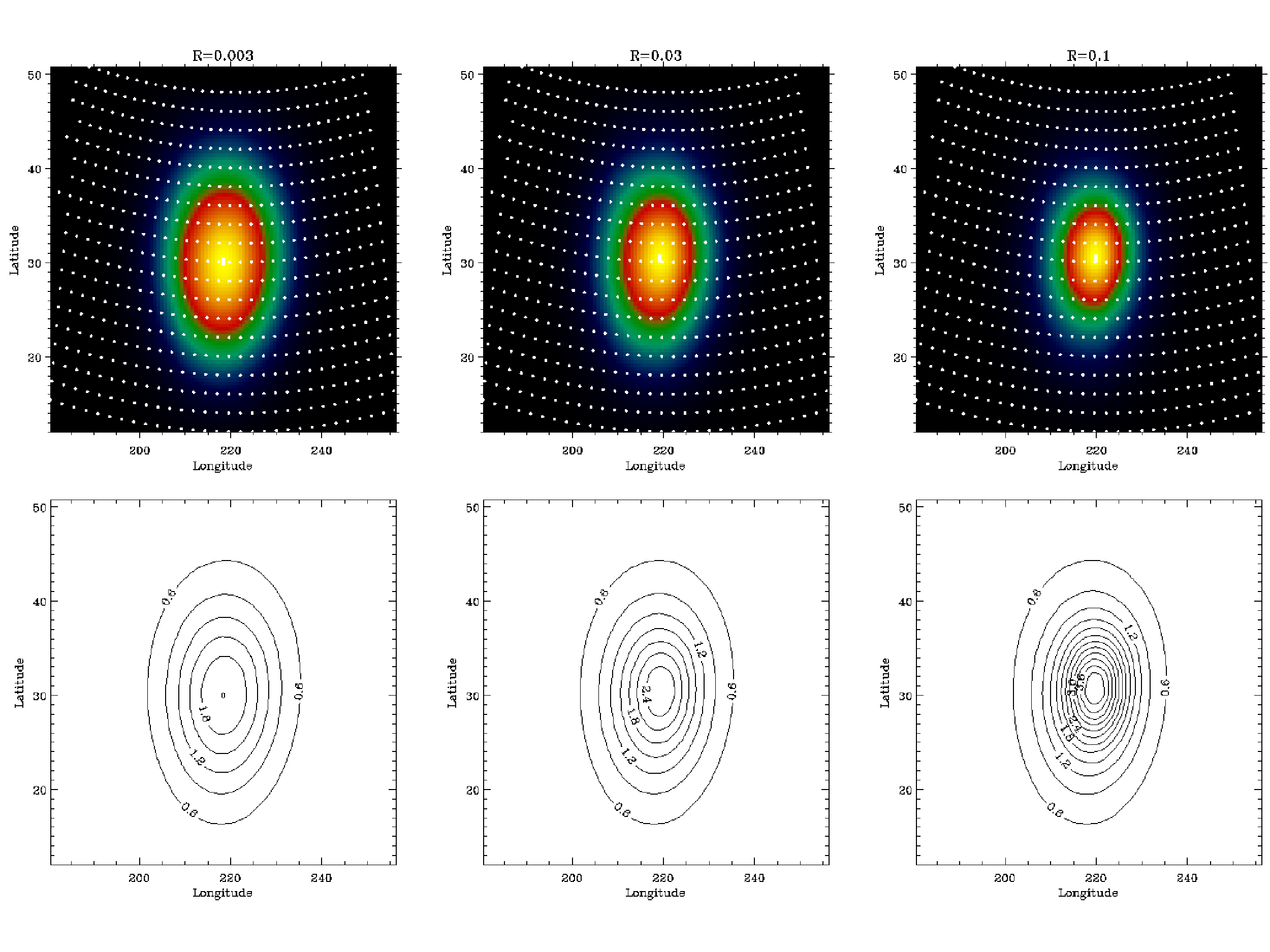}{3.3in}{0}{55}{55}{-200}{-20}
\caption{Simulations of an actual {\em Ulysses}/GAS particle beam map from
  2001 December 12 (in ecliptic coordinates), shown in both image (top)
  and contour plot (bottom) formats (Wood et al.\ 2015b).  The maps are
  generated assuming the best-fit He flow vector from Wood et al.\ (2015a)
  and a temperature of $T=8500$~K.  The color scale in the images is
  rescaled in each panel so that the maximum is always bright yellow.
  Maps are computed assuming three different degrees of
  contamination from O and Ne, corresponding to $R=0.003$ (left),
  $R=0.03$ (middle), and $R=0.1$ (right).}
\end{figure}
\begin{figure}
\plotfiddle{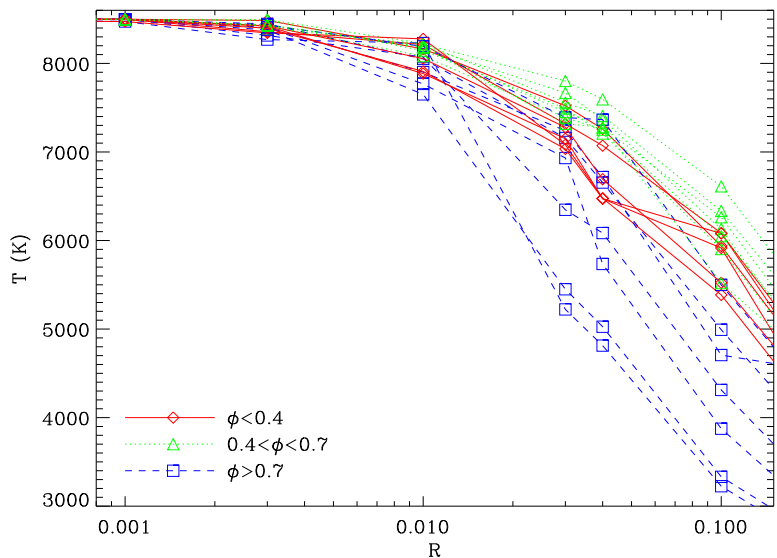}{2.2in}{90}{45}{45}{180}{-40}
\caption{Single map fits have been performed on synthetic {\em Ulysses}
  particle beam maps, computed assuming various values of R (measuring
  contamination of O and Ne flux), with map sampling based on 20 real
  observations (Wood et al.\ 2015b).  For each of the 20 cases, the
  measured temperature is plotted versus R, with different symbols and
  line styles indicating different ranges of orbital phase, $\phi$.}
\end{figure}

     The question is then whether $R\approx 0.03$ is low enough
to be a plausible value.  This is about an order of magnitude higher
than the $R=0.0025$ value estimated above based on our best estimates
of the quantities in equation (1).  Thus, we conclude that heavy
element contamination of {\em Ulysses} data is unlikely to be the
primary cause of the ISM temperature discrepancy between {\em Ulysses}
and IBEX (Wood et al.\ 2015b).

\section{The Effects of Velocity Distribution Asymmetries on IBEX Data}

     The IBEX spacecraft is designed to produce a sky map of
energetic neutral atom fluxes once every 6 months.  In its highly
elliptical orbit around Earth, the spinning spacecraft allows its
detectors to scan a roughly $14^{\circ}$ wide swath of sky in a
given orbit.  The spin axis is kept pointed roughly towards the
Sun, meaning data from each individual IBEX orbit provides a
plot of count rate versus ecliptic latitude.  After 7-8 days the
orbit is changed so that the new scan plane is rotated typically
$7^{\circ}-8^{\circ}$ in longitude.  In this way, IBEX gradually
covers the entire sky every 6 months.

\begin{figure}[t]
\plotfiddle{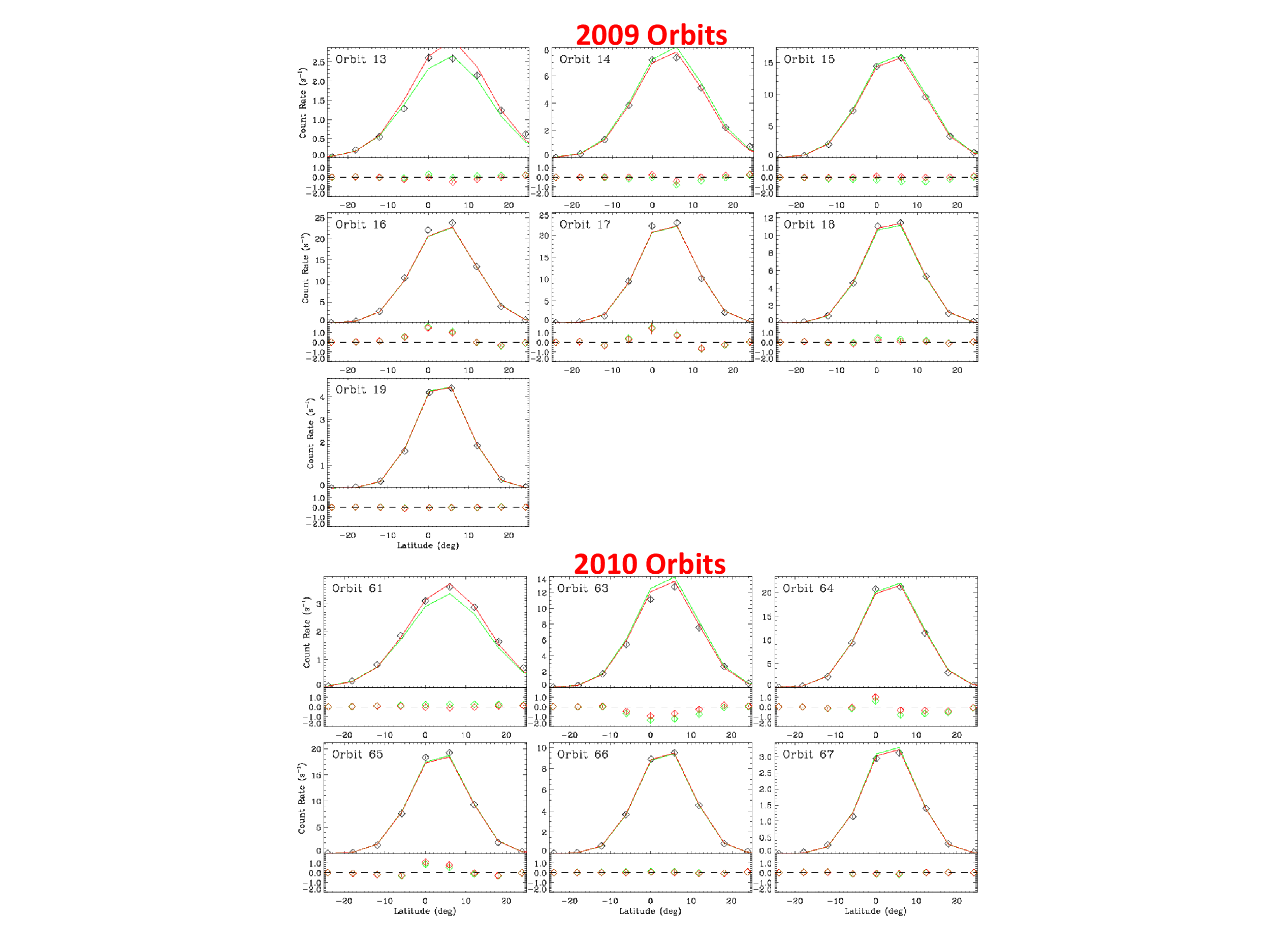}{6.0in}{0}{95}{95}{-350}{-30}
\caption{The first two years of IBEX observations of ISM neutral He.
  Each panel is a plot of count rate versus ecliptic latitude for a
  given orbit, with each orbit representing a different longitudinal
  plane.  Green lines are a fit to the data assuming that the
  source distribution is a simple Maxwellian far from the Sun, while
  red lines are a fit assuming the distribution far from the Sun is the
  sum of two Maxwellians, with
  a separation between the two Maxwellians of $\Delta V=0.4$ km~s$^{-1}$.
  Residuals are shown below each panel for both fits.}
\end{figure}
     The interstellar neutrals are only observable at one time each year,
with peak fluxes from early January to late February (M\"{o}bius et al.\
2012).  Figure~3 shows the first two years of IBEX observations
of the He beam.  These are the same data analyzed by
M\"{o}bius et al.\ (2012) and Bzowski et al.\ (2012) in the first
assessments of the He flow vector from IBEX.
We have fitted these data with an analysis approach essentially
identical to that used to analyze {\em Ulysses} data (Wood et al.\
2015a).  Our best fit is shown in Figure~3.

\begin{figure}[t]
\plotfiddle{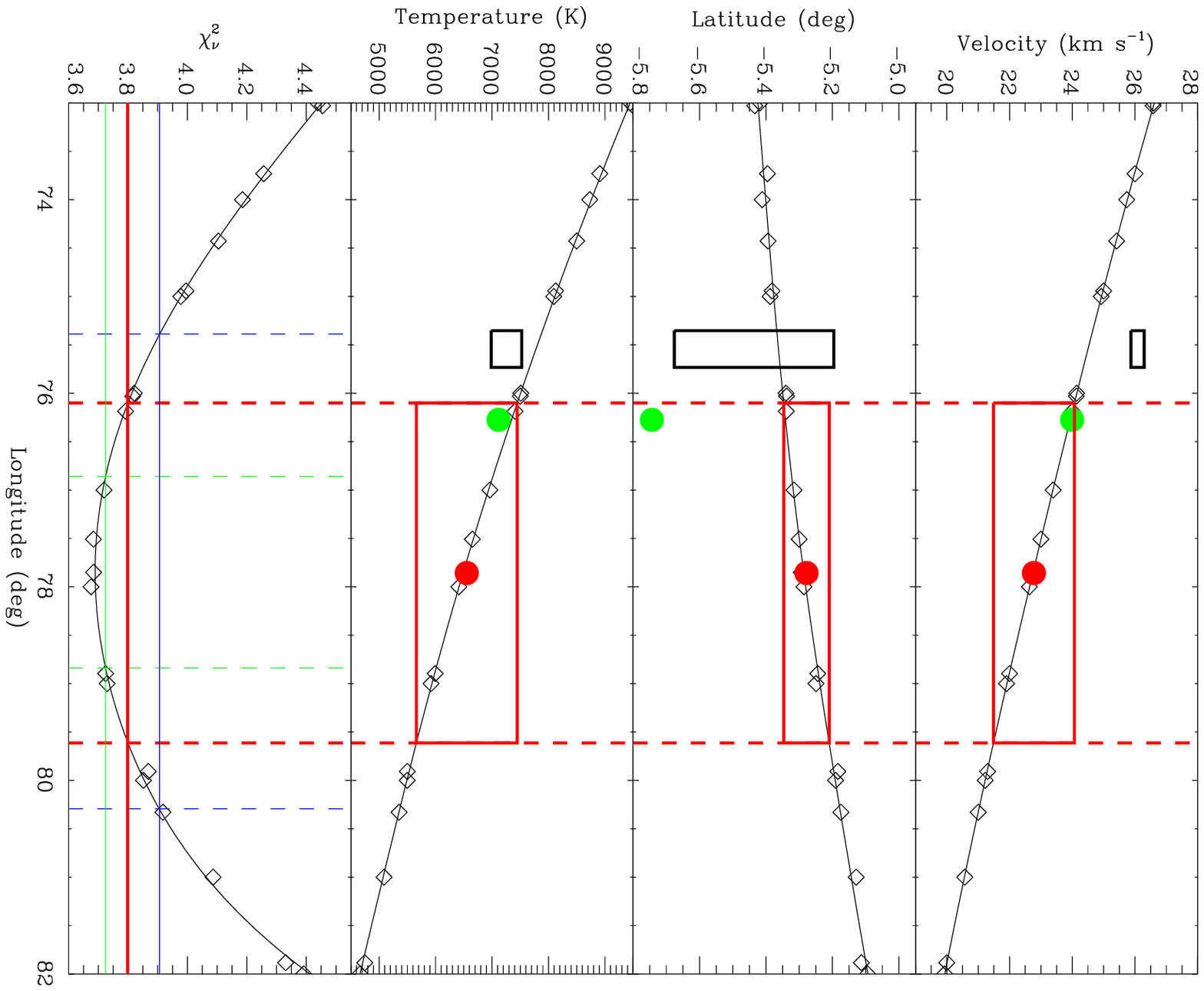}{5.5in}{90}{90}{90}{360}{-80}
\caption{The diamonds show the results of various fits to the first two
  years of IBEX observations of the ISM He flow, assuming a simple
  Maxwellian distribution far from the Sun, with one of the four
  fit parameters held constant in each fit.  The upper three panels plot
  ISM velocity, flow latitude, and temperature versus flow longitude,
  and the bottom panel plots $\chi^2$ versus flow longitude.  Horizontal
  lines in the bottom panel indicate 1, 3, and 5 $\sigma$ confidence
  levels.  The 3$\sigma$ level is used to define the IBEX error boxes in
  red in the upper 3 panels, which can be compared with the {\em Ulysses}
  error boxes in black, from Wood et al.\ (2015a).  Red circles indicate
  the best single-Maxwellian IBEX fit, while the green circles indicate
  the best fit assuming the He flow far from the Sun is the sum
  of two closely-spaced Maxwellians.}
\end{figure}
     The four free parameters
of the fit are flow velocity ($V_{ISM}$), temperature ($T$), flow
longitude ($\lambda$), and flow latitude ($\beta$).  As shown in
Figure~4, the four fit parameters are highly correlated, so you have
long tubes in four-dimensional chi-squared space where the fits
are reasonably good (Bzowski et al.\ 2012; M\"{o}bius et al.\ 2012;
McComas et al.\ 2012).  Nevertheless, the bottom panel of Figure~4
shows that there is a well-defined $\chi^2$ minimum, allowing clear
best-fit values and error bars to be computed.  Following the
example of Wood et al.\ (2015a) in the analysis of {\em Ulysses}
data, we use 3$\sigma$ confidence intervals to define the error
bounds, leading to the red error boxes in the figure, corresponding
to the following values:  $V_{ISM}=22.8\pm 1.3$ km~s$^{-1}$,
$\lambda=77.9\pm 1.8^{\circ}$, $\beta=-5.3\pm 0.1^{\circ}$~deg,
and $T=6550\pm 890$~K.

     These values are basically in agreement with the best-fit
values reported by Bzowski et al.\ (2012), who analyzed these
data with a similar forward modeling approach.  They can also be
compared with our {\em Ulysses} measurements
($V_{ISM}=26.08\pm 0.21$ km~s$^{-1}$, $\lambda=75.54\pm 0.19^{\circ}$,
$\beta=-5.44\pm 0.24^{\circ}$, and $T=7260\pm 270$~K), which are shown
as black boxes in Figure~4.  The apparent inconsistency between
the {\em Ulysses} and IBEX measurements was initially an issue of
much debate, though consideration of more recent observations from
IBEX seems to have yielded IBEX measurements more consistent with
those of {\em Ulysses} (McComas et al.\ 2015; Leonard et al.\ 2015;
M\"{o}bius et al.\ 2015), with the possible exception of
temperature, as discussed in Section~2.  The cause of the discrepancy
with the first two years of IBEX data seen in Figure~4 has
not been explained.  However, we note here that if we used 5$\sigma$
error boxes instead of 3$\sigma$, the {\em Ulysses} and IBEX boxes
would overlap, except for $V_{ISM}$.  In any case, our analysis
combined with that of Wood et al.\ (2015a) suggests the {\em Ulysses}
analysis provides more precise measurements of the flow vector than
the IBEX analysis, as indicated by the smaller size of the
{\em Ulysses} boxes relative to IBEX.  The reason for this lies in
the ability of {\em Ulysses} to observe the He beam from different
locations, above and below the ecliptic.  This breaks the parameter
degeneracies that bedevil IBEX, which can only observe the He beam
from one location (Wood et al.\ 2015a).

     The analysis of IBEX data just presented relies on the usual
assumption that the ISM He distribution far from the Sun is a
simple Maxwellian.  The best fit in Figure~3 has $\chi^2=3.7$,
well above the ideal value of $\chi^2\approx 1$,
indicating that there is significant room for improvement with
regards to how well we are fitting the IBEX data.  Perhaps
inaccuracies in the single Maxwellian assumption may be at
least partly responsible for this.  A non-Maxwellian distribution
could also potentially affect the best-fit He flow parameters
that we have derived.  Solar wind distributions are often
found to have significant asymmetries (Marsch et al.\ 1982),
so perhaps ISM distributions might as well.

     In order to explore the effects of a non-Maxwellian
distribution on the IBEX analysis, we experiment with a
distribution that is the sum of two closely spaced identical
Maxwellians, oriented along the ISM magnetic field line.
This is meant to approximate a bi-Maxwellian distribution
with $T_{\parallel}>T_{\perp}$.  We assume the ISM field is
oriented as suggested by the IBEX ribbon,
towards ($\lambda$,$\beta$)=($221^{\circ}$,$39^{\circ}$)
(Funsten et al.\ 2009).
Fits are then performed to the IBEX data assuming different
velocity separations for the two Maxwellians.  Figure~5
shows how $\chi^2$ varies with assumed separation.  The
$\chi^2$ drops from 3.7 to 2.7 when the separation is
increased from $\Delta V=0$ to $\Delta V=0.4$ km~s$^{-1}$,
a surprisingly large amount considering just how small an
asymmetry this shift induces in the overall distribution.
For comparison, the thermal width of the He distribution for
an 8000~K He gas is 5.7 km~s$^{-1}$.  However, Figure~5 also
shows that $\chi^2$ starts increasing for even larger
separations (i.e., larger asymmetries).
\begin{figure}[t]
\plotfiddle{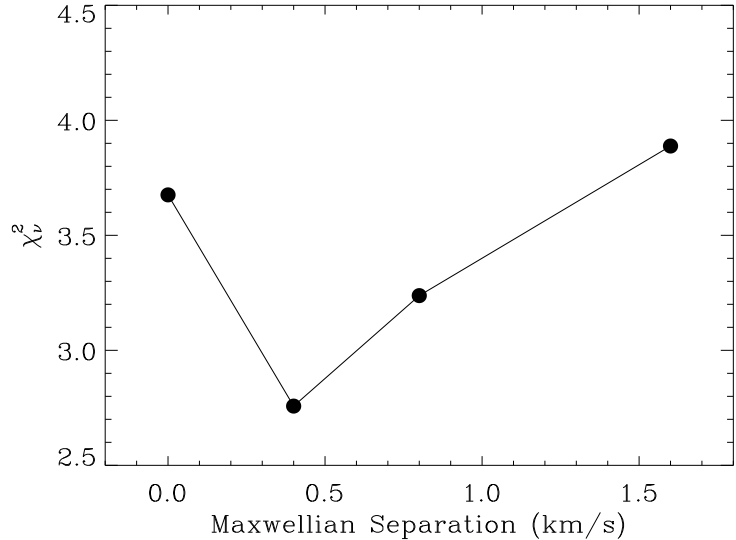}{2.4in}{90}{45}{45}{180}{-40}
\caption{The $\chi^2$ statistic is plotted versus Maxwellian
  velocity separation, for fits to the first two years of
  IBEX data assuming the He flow far from the Sun can be
  represented as the sum of two closely-spaced Maxwellians.}
\end{figure}

     Figure~3 shows the fit to the data for the $\Delta V=0.4$
km~s$^{-1}$ fit, and Figure~4 shows the new best-fit He flow
parameters that result from the $\Delta V=0.4$ km~s$^{-1}$
assumption ($V_{ISM}=24.0$ km~s$^{-1}$, $\lambda=76.3^{\circ}$,
$\beta=-5.7^{\circ}$, and $T=7120$~K).  We find a lack of robustness
for fits assuming these two-Maxwellian distributions, meaning such
fits do not have as clean and smooth a $\chi^2$ curve as seen in the
bottom panel of Figure~4 for the simple Maxwellian fit.  The cause of
this is uncertain, but as a consequence we do not try to quote error
bars for the two-Maxwellian fits at this time.  Nevertheless, it is
interesting to note from Figure~4 that the best-fit parameters have
changed significantly for the $\Delta V=0.4$ km~s$^{-1}$ fit, and they
have actually moved conveniently in the direction of the {\em Ulysses}
values.

     We conclude from this excercise that small asymmetries
in the ISM velocity distribution can potentially affect both
the quality of fit and the best-fit parameters in analyses
of IBEX data.  It is therefore worthwhile to explore further
the possible effects of non-Maxwellian distributions, which
may influence the consistency of parameters derived by
IBEX and {\em Ulysses}.  Future work should involve actual
bi-Maxwellian distributions rather than the two-Maxwellian
summation approximation.

\ack

This work has been supported by NASA award NNH13AV19I to
the Naval Research Laboratory.

\section*{References}
\begin{thereferences}
\item Banaszkiewicz, M., Witte, M., \& Rosenbauer, H. 1996, A\&AS, 120, 587
\item Bzowski, M., et al. 2012, ApJS, 198, 12
\item Bzowski, M., Sok\'{o}\l, J. M., Kubiak, M. A., \& Kucharek, H. 2013,
  A\&A, 557, A50
\item Bzowski, M., Kubiak, M. A., H{\l}ond, M., Sok\'{o}{\l}, J. M.,
  Banaszkiewicz, M., \& Witte, M. 2014, A\&A, 569, A8
\item Funsten, H. O., et al. 2009, Science, 326, 964
\item Leonard, T. W., et al. 2015, ApJ, 804, 42
\item Marsch, E., M\"{u}hlh\"{a}user, K. -H., Schwenn, R., Rosenbauer, H.,
  Pilipp, W., \& Neubauer, F. M. 1982, J.~Geophys.~Res., 87, 52
\item McComas, D. J., et al. 2012, Science, 336, 1291
\item McComas, D. J., et al. 2015, ApJ, 801, 28
\item M\"{o}bius, E., et al. 2012, ApJS, 198, 11
\item M\"{o}bius, E., et al. 2015, ApJS, submitted
\item Schwadron, N., et al. 2015, ApJS, submitted
\item Slavin, J. D., \& Frisch, P. C. 2008, A\&A, 491, 53
\item Witte, M. 2004, A\&A, 426, 835
\item Wood, B. E., M\"{u}ller, H. -R., Bzowski, M., Sok\'{o}\l, J. M.,
  M\"{o}bius, E., Witte, M., \& McComas, D. J. 2015b, ApJS, in press
\item Wood, B. E., M\"{u}ller, H. -R., \& Witte, M. 2015a, ApJ, 801, 62
\item Yamamura, Y., \& Tawara, H. 1996, Atomic Data and Nuclear Data
 Tables, 62, 149
\end{thereferences}

\end{document}